# On the correlation properties of thermal noise in fluids

Vladimír Lisý, Jana Tóthová

*Department of Physics, Technical University of Košice,*
*Park Komenského 2, 04200 Košice, Slovakia*

Lukáš Glod

*Department of Mathematics and Physics, University of Security Management,*
*Kukučínova 17, 04001 Košice, Slovakia*

**Abstract** The properties of the thermal force driving micron particles in incompressible fluids are studied within the hydrodynamic theory of the Brownian motion. It is shown that the assumption used for the hydrodynamic Langevin equation in its usual form with the initial time $t = 0$, according to which the random force at a time $t > 0$ and the velocity of the particle at $t = 0$ are uncorrelated, leads to super-diffusion of the particle. To obtain the correct Einstein diffusion at long times, the mentioned hypothesis must be abandoned, which however does not contradict causality. The corresponding correlations are explicitly evaluated. We consider also the "color" of thermal noise, recently measured experimentally [Th. Franosch *et al.*, Nature 478, 85 (2011)], and correct the interpretation of these experiments. The time correlation functions for the thermal random force are obtained using the exact solution of the Langevin equation, and on the basis of the theorem that in the linear response theory connects the mobility of the particle and its velocity autocorrelation function.

## 1 Introduction

It has been known for a long time [1, 2] that the Langevin equation describing the Brownian motion of particles is valid only under limited conditions: the density of the particles must be much larger than the density of the surroundings (simultaneously, the observation times must be small). Alternatively, the time of observation must be long. However, then the particles are in steady motion and the Einstein theory of the Brownian motion is appropriate [3, 4]. When the density of the particles is comparable to that of the fluid, the standard Einstein-Langevin theory of the Brownian motion should be significantly changed: instead of the Stokes force modeling the friction that the Brownian particle feels during its motion, a more correct resistance force should be used. It must reflect the history of the particle dynamics, i.e., at time $t$ it should depend on the particle state of the motion in all the preceding moments of times. Assuming incompressible fluids, this "history force" is the Boussinesq-Basset force [5, 6]. It has been demonstrated experimentally that the modified Langevin equation very well describes the Brownian motion in fluids, both in situations when a particle is free and when it is confined in an external potential well [7 - 10] (see also the commentary [11]). At very short times the mean square displacement of a particle increases as $t^2$, which corresponds to ballistic motion. At long times, hydrodynamic vortices in the liquid created by the particle's motion lead to memory effects. These effects are displayed in the time correlation functions, such as the velocity autocorrelation function that decays much more slowly than exponentially, exhibiting "long-time tails" ($t^{-3/2}$ and $t^{-7/2}$ tails for a free particle [1, 12, 13] and a particle in a harmonic trap [14, 15], respectively). The Langevin equation contains a random force that accounts for the collisions with the surrounding molecules. This thermal force is assumed to be "white", which means that it is delta-correlated in time and its frequency spectrum is constant. This is true until the friction force in the equation is well described by the Stokes force, proportional to the particle velocity. In the hydrodynamic approach with the Boussinesq friction, the thermal force is no longer the white noise. Due to the fluctuation-dissipation theorem its values at different moments of time correlate. The present work is devoted just to the study of the correlation properties of this colored thermal force that drives the motion of Brownian particles in fluids. It is shown that when the commonly accepted properties of the random force in the hydrodynamic Langevin equation for incompressible fluids are assumed (if the initial time is taken at $t = 0$), this results in an unexpected super-diffusive motion of the particles. Since normal diffusion should take place, which is an undoubted experimental fact, this contradiction needs to be resolved. We show that the expected Einstein diffusion at long times requires that the time correlation function of the random force at a moment $t > 0$ and the particle velocity at $t = 0$ must be nonzero. This apparent paradox is explained. We then calculate the correlation function of the thermal force itself and show that its properties (the color) significantly differ from those found in the literature [10, 16, 17]. Although the thermal force in the Langevin equation is a physical reality and should be observable [18], its color has been directly measured only very recently in the work by Franosch et al. [10]. Combining strong optical trapping with high-resolution interferometric detection, the correlations in thermal noise became directly accessible by calculating the positional autocorrelation function from the recorded position fluctuations of the particle trapped in a harmonic potential. Analyzing the hydrodynamic Langevin equation, the time correlation function for the noise was found proportional to $t^{-3/2}$. We show that along with the $t^{-3/2}$ term obtained in [10], this correlation function contains a more slowly decaying term that depends on the time as $t^{-1/2}$.

## 2 Hydrodynamic Brownian motion in fluids

As mentioned in Introduction, the standard Langevin equation for Brownian particles in fluids should be changed so that instead of the Stokes friction force it contains the Boussinesq force. It has the form [5, 6]

$$F_B(t) = -\gamma \left\{ \upsilon(t) + \frac{\rho R^2}{9\eta} \frac{d\upsilon}{dt} + \sqrt{\frac{\rho R^2}{\pi \eta}} \int_{-\infty}^{t} \frac{d\upsilon}{dt'} \frac{dt'}{\sqrt{t-t'}} \right\} \quad (1)$$

that follows from the non-stationary Navier-Stokes equations of motion for incompressible fluids [19]. In this equation $\upsilon = dx/dt$ is the particle velocity in a chosen direction, e.g., $x$, $\gamma =$ $6\pi\eta R$ is the Stokes friction coefficient for spherical particles with radius $R$, $\rho$ is the density and $\eta$ is the dynamic viscosity of the solvent. The Boussinesq force introduces the viscous after-effect and consequently the memory in the particle dynamics. The loss of this hydrodynamic memory is characterized by the vorticity time $\tau_R = R^2\rho/\eta$. The generalized Langevin equation with the force (1) in the form used in the literature [12-17], is (see, however, the next section)

$$M\dot{\upsilon}(t) + \gamma\upsilon(t) + \int_0^t \Gamma(t-t')\dot{\upsilon}(t')\,dt' = F + \zeta(t). \quad (2)$$

Here, the random noise force $\zeta(t)$ with zero mean drives the particles of mass $M_p$ ($M = M_p + M_s/2$ with $M_s$ being the mass of the solvent displaced by the particle), and $F$ is an external force. The kernel $\Gamma$ is $\Gamma(t) = \gamma(\tau_R/\pi t)^{1/2}$. The usual Brownian relaxation time $\tau = M/\gamma$ is connected to $\tau_R$ by the relation $\tau_R/\tau = 9\rho/(2\rho_p + \rho)$, where $\rho_p$ is the density of the particle. Equation (2) for $F = 0$ describes the zero-mean fluctuations $\upsilon(t)$. Now our aim is to calculate the velocity autocorrelation function (VAF) of the free particle, $\phi(t) = \langle \upsilon(t)\upsilon(0)\rangle$, and its mean square displacement (MSD) for the motion along the direction $x$, $X(t) = \langle \Delta x^2(t)\rangle = \langle [x(t) - x(0)]^2 \rangle$. The brackets $\langle \ldots \rangle$ remain for statistical averaging. Since the quantities $\upsilon(t)$ and $x(t)$ are stochastic variables, we do not use zero initial conditions, as it is usually done for them [12-14, 16, 17]. Instead, assuming the initial equilibrium between the particle and the solvent, in agreement with the equipartition theorem for a particle of mass $M$, the condition $\phi(0) = k_B T/M$ is used for the VAF. Then, multiplying Eq. (2) by $\upsilon(0)$ and assuming that $\langle \zeta(t)\upsilon(0)\rangle = 0$, after the average one obtains

$$M\dot{\phi} + \gamma\phi + \int_0^t \Gamma(t-t')\dot{\phi}(t')\,dt' = 0. \quad (3)$$

In the Laplace transformation the solution for $\tilde{\phi}(s) = \mathcal{L}\{\phi(t)\}$ is

$$\tilde{\phi}(s) = \frac{k_B T}{M}\frac{M + \tilde{\Gamma}(s)}{\gamma + s\left[M + \tilde{\Gamma}(s)\right]}, \quad (4)$$

with $\tilde{\Gamma}(s) = \gamma\tau_R^{1/2}s^{-1/2}$. The inverse transform of (4) is found after expanding this expression in simple fractions. If $\lambda_{1,2} = -(\tau_R^{1/2}/2\tau)(1 \mp \sqrt{1 - 4\tau/\tau_R})$ are the roots of the equation $s + (\tau_R s)^{1/2}\tau^{-1} + \tau^{-1} = 0$, we have

$$\tilde{\phi}(s) = \frac{k_B T}{Ms^{1/2}}\left(\frac{C_1}{s^{1/2} - \lambda_1} + \frac{C_2}{s^{1/2} - \lambda_2}\right), \quad (5)$$

where $C_1 = (\lambda_1 + \tau_R^{1/2}\tau^{-1})/(\lambda_1 - \lambda_2)$, and $C_2$ is obtained by exchanging $\lambda_1 \rightleftarrows \lambda_2$. Using the inverse transform $\mathcal{L}^{-1}\{s^{-1/2}(s^{1/2} - \lambda)^{-1}\} = \exp(\lambda^2 t)\text{erfc}(-\lambda\sqrt{t})$ [20], where erfc is the complementary error function, one finds

$$\phi(t) = \frac{k_B T}{M} \sum_{i=1}^{2} C_i \exp(\lambda_i^2 t) \operatorname{erfc}(-\lambda_i \sqrt{t}). \qquad (6)$$

The asymptotes of this expression are

$$\phi(t) \approx \frac{k_B T}{M}(1 - t/\tau + ...), \quad t \to 0, \qquad (7)$$

$$\phi(t) \approx \frac{k_B T}{M}\left(\frac{\tau_R}{\pi t}\right)^{1/2}\left[1 - \frac{\tau_R}{2t}\left(1 - \frac{2\tau}{\tau_R}\right) + ...\right], \quad t \to \infty. \qquad (8)$$

Representing the distance a particle moves in time as an integral of its velocity, $x(t) - x(0) = \int_0^t \upsilon(s)\,\mathrm{d}s$, the MSD is obtained as $X(t) = 2\int_0^t (t-s)\phi(s)\,\mathrm{d}s$ [21],

$$X(t) = \frac{2k_B T}{M}\frac{1}{\lambda_1 - \lambda_2}\left\{\frac{\lambda_2}{\lambda_1^4}\left[\frac{4}{3\sqrt{\pi}}(\lambda_1 t^{1/2})^3 + \lambda_1^2 t \right.\right.$$
$$\left.\left. + \frac{2\lambda_1}{\sqrt{\pi}} t^{1/2} + 1 - \exp(\lambda_1^2 t)\operatorname{erfc}(-\lambda_1 t^{1/2})\right] - (\lambda_1 \rightleftarrows \lambda_2)\right\}. \qquad (9)$$

At short times it shows the ballistic behavior, $X(t \to 0) \sim k_B T t^2 / M$, which is confirmed experimentally [8 - 11]. At $t \to \infty$, up to the first term decreasing with time, we have

$$X(t) = \frac{2k_B T}{M}\left[\frac{4}{3}\left(\frac{\tau_R t^3}{\pi}\right)^{1/2} + (\tau - \tau_R)t + 2\left(1 - \frac{2\tau}{\tau_R}\right)\left(\frac{\tau_R^3 t}{\pi}\right)^{1/2}\right.$$
$$\left. -(\tau^2 - 3\tau\tau_R + \tau_R^2) + (\tau_R - 3\tau)(\tau_R - \tau)\left(\frac{\tau_R}{\pi t}\right)^{1/2} + ...\right]. \qquad (10)$$

Since $X(t \to \infty) \sim t^{3/2}$, the solution has a super-diffusive character. Of course, this unexpected result is not correct. It contradicts both the Einstein-Langevin theory and numerous experiments. Equation (6) and the following equations (8) – (10) significantly differ from those found in the literature for the VAF as a solution of Eq. (2) with $F = 0$ [1, 2, 12 - 17],

$$\phi(t) = \frac{k_B T}{M}\frac{1}{\lambda_1 - \lambda_2}\left[\lambda_1 \exp(\lambda_2^2 t)\operatorname{erfc}(-\lambda_2 \sqrt{t}) - \lambda_2 \exp(\lambda_1^2 t)\operatorname{erfc}(-\lambda_1 \sqrt{t})\right], \qquad (11)$$

which looks quite similar to (6), but its asymptote is very different from Eq. (8):

$$\phi(t) \approx \frac{k_B T}{2M}\frac{\tau \tau_R^{1/2}}{\pi^{1/2} t^{3/2}}\left[1 - \frac{3}{2}\left(1 - \frac{2\tau}{\tau_R}\right)\frac{\tau_R}{t} + ...\right], \quad t \to \infty. \qquad (12)$$

The MSD that follows from Eq. (11) is

$$X(t) = 2D\left\{t - 2\left(\frac{\tau_R t}{\pi}\right)^{1/2} + \tau_R - \tau\right.$$

$$\left. + \frac{1}{\tau}\frac{1}{\lambda_2 - \lambda_1}\left[\frac{\exp(\lambda_2^2 t)}{\lambda_2^3}\text{erfc}\left(-\lambda_2\sqrt{t}\right) - \frac{\exp(\lambda_1^2 t)}{\lambda_1^3}\text{erfc}\left(-\lambda_1\sqrt{t}\right)\right]\right\}, \quad (13)$$

with the $t \to \infty$ behavior

$$X(t) \approx 2Dt\left\{1 - 2\left(\frac{\tau_R}{\pi t}\right)^{\frac{1}{2}} + \frac{2}{9}\left(4 - \frac{M_p}{M_s}\right)\frac{\tau_R}{t} - \frac{1}{9\sqrt{\pi}}\left(7 - 4\frac{M_p}{M_s}\right)\left(\frac{\tau_R}{t}\right)^{\frac{3}{2}} + ...\right\}, \quad (14)$$

where $D = k_B T \tau / M$ is the Einstein diffusion coefficient of the particle. Figures 1 and 2 show the difference between the solutions (6) and (11) for the VAF, and (9) and (13) for the MSD, respectively, for Brownian particles studied in the experiments [9]. These solutions are also compared to the results of the standard Langevin theory [4]. Equations (10) and (14) coincide at $\tau_R \to 0$ but for particles with $\rho_p$ close to $\rho$ the difference between them is significant. The question arises how to get the correct solutions (11) and (13).

In the cited papers [10, 12-14, 16, 17], the zero initial condition for the particle velocity is assumed. For example, in Ref. [13] both the projection of the particle position vector on the axis $x$, $x(t)$, and the velocity, $\upsilon(t)$, are assumed zero at $t = 0$: $x(0) = \upsilon(0) = 0$. It is argued that within the approximation of small thermal displacements there is no loss of generality in choosing these trivial conditions. Similarly, one finds in Ref. [14] that the velocity of the particle is determined by its velocity at earlier times via backflow effects in the fluid, but it is assumed that the particle is at the equilibrium position $x(0) = 0$ and at rest for $t \leq 0$. The latter condition is used also in the work [12], where independently the same solution for the VAF and MSD as in [13] has been obtained coming from the linear response theory [22]. In our opinion, the simultaneous use of the conditions $x(0) = 0$ and $\upsilon(0) = 0$ is not consistent for random variables $x(t)$ and $\upsilon(t)$. Moreover, it is either assumed or it follows from the calculations in all the works that the equipartition theorem holds, i.e. that $\langle \upsilon^2(0) \rangle$ is nonzero. We will show that if the above assumptions are abandoned, the correct (normal) diffusion can be obtained only if the random force at $t > 0$ is correlated with the particle velocity at $t = 0$. That is, we must require for Eq. (2) that $\langle \zeta(t)\upsilon(0) \rangle$ is nonzero at $t > 0$. This will be shown in the next section. Note that a different method for finding the MSD of the non-Markovian motion of a spherical Brownian particle in an infinite medium has been proposed in Ref. [23]. Using the Laplace transform, Eq. (2) was directly solved for the coordinate of the particle, $x(t)$. In the time domain this solution is expressed as a convolution of a function $G(t)$, known from the Langevin equation, and the random force $\zeta(t)$. Then the function $G$ is used to construct characteristic multidimensional functions $g_L(\lambda; t)$ of this integral transform. The one-dimensional function $g_1(\lambda; t)$ determines the MSD according to the formula $X(t) = \partial^2 g_1(\lambda;t)/\partial(i\lambda)^2$ at $\lambda = 0$. For the limits considered in [23] the results agree with the above given solutions for $X(t)$; in particular, at short times the motion is ballistic and the long-time limit corresponds to the solution (14). However, the properties of the thermal force were not discussed in [23].

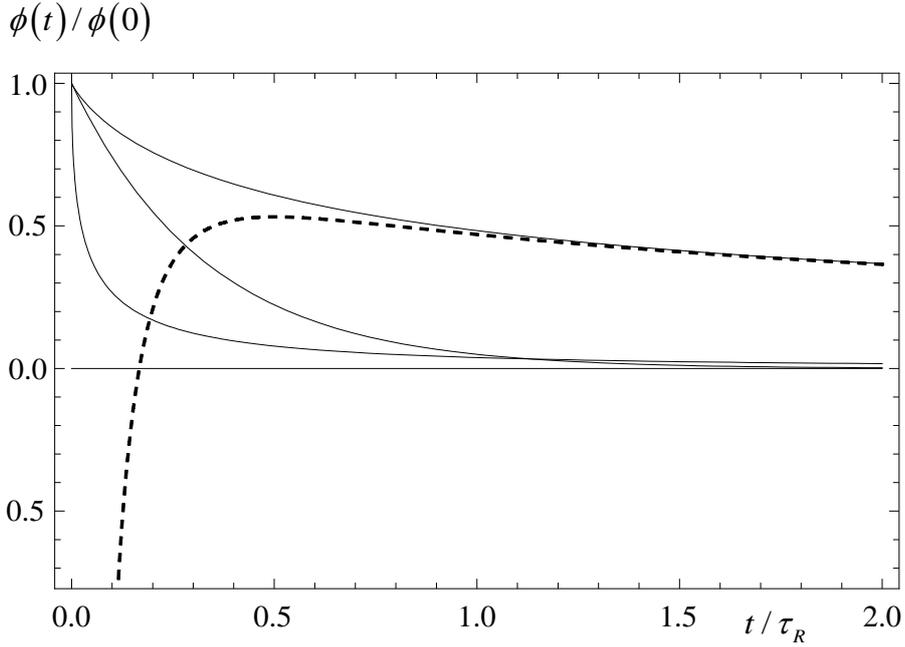

**Fig. 1** Normalized velocity autocorrelation function for resin particles (with $R = 1.25$ μm and the density $1.5 \cdot 10^3$ kg·m$^{-3}$) in water at room temperature ($\eta = 10^{-3}$ Pa·s, $\rho = 10^3$ kg·m$^{-3}$, $\tau_R = 1.56 \cdot 10^{-6}$ s, $\tau = 0.52 \cdot 10^{-6}$ s). The upper line corresponds to Eq. (6). The lower full lines are for the standard Langevin theory, the hydrodynamic theory of the Brownian motion, Eq. (11), and the Einstein limit ($\phi = 0$), respectively [4]. The dashed line is for the long-time limit (8) that after unphysical values at short times very early joins the exact solution (6)

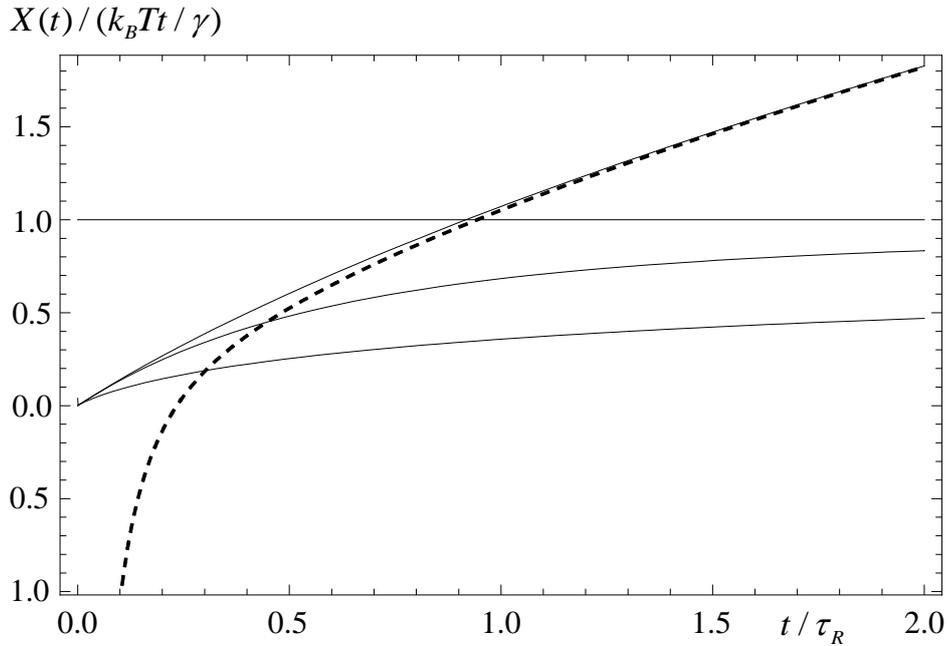

**Fig. 2** Mean square displacement for resin particles in water, normalized to the Einstein result. The full line corresponds to super-diffusion from Eq. (9). The lower full lines are for the standard Langevin theory [4] and the hydrodynamic theory (13), and the dashed line is for long-time limit (10). The numerical parameters are the same as in Fig. 1

# 3 Properties of thermal random force in incompressible fluids

In spite of the abovementioned inconsistencies the solutions (11) and (13) [12 - 14] correctly describe the expected Einstein diffusion at long times for any relation between the densities $\rho_p$ and $\rho$. This problem could be resolved as follows. If we assume that the studied process is in thermodynamic equilibrium, the initial value $\upsilon(0)$ (for which we assume that the equipartition holds) should be the result of the long time memory in the system. The correct Langevin equation to be used in the description of the hydrodynamic Brownian motion should have the form

$$M\dot{\upsilon}(t)+\gamma\upsilon(t)+\int_{-\infty}^{t}\Gamma(t-t')\dot{\upsilon}(t')\,\mathrm{d}t'=\eta(t), \quad (15)$$

where $\zeta(t)$ from Eq. (2) is $\zeta(t)=\eta(t)-\int_{-\infty}^{0}\Gamma(-t')\dot{\upsilon}(t')\,\mathrm{d}t'$. While $\langle\eta(t)\upsilon(0)\rangle = 0$ due to causality, now we can assume that the force $\zeta(t)$ and the velocity $\upsilon(0)$ correlate. Let the correlator $\langle\zeta(t)\upsilon(0)\rangle = Z(t)$ with the Laplace transform $\tilde{Z}(s)$. In the same way as before, one finds the equation for the VAF,

$$M\dot{\phi}+\gamma\phi+\int_{0}^{t}\Gamma(t-t')\dot{\phi}(t')\,\mathrm{d}t'=Z(t). \quad (16)$$

Its solution in the Laplace transform is

$$\tilde{\phi}(s)=\frac{k_{B}T}{M}\frac{1+(\tau_{R}/s)^{1/2}\tau^{-1}+\tilde{Z}(s)/k_{B}T}{s+(\tau_{R}s)^{1/2}\tau^{-1}+\tau^{-1}}. \quad (17)$$

Let us require that in the long time limit the particle is in the diffusion regime with the Einstein's MSD proportional to $t$. This may happen only if $\tilde{\phi}(s)$ at $s \to 0$ tends to a constant, namely, $\tilde{\phi}(s) \to k_{B}T/\gamma = k_{B}T\tau/M$. Consequently, $\tilde{Z}(s)/k_{B}T \approx -(\tau_{R}/\tau^{2}s)^{1/2}$ for small $s$. Other terms on the right that could go to zero if $s \to 0$ must be excluded since at $s \to \infty$ there should be $s\tilde{\phi}(s) \approx k_{B}T/M$ to have the right $t = 0$ limit. The relation $\tilde{Z}(s)/k_{B}T = -(\tau_{R}/\tau^{2}s)^{1/2}$ thus holds for every $s$, so that the VAF is

$$\tilde{\phi}(s)=\frac{k_{B}T}{M}\frac{1}{s+(\tau_{R}s)^{1/2}\tau^{-1}+\tau^{-1}}, \quad (18)$$

which corresponds to the solution (11). Note that if it is only required more generally that $\phi(t) \to 0$ as $t \to \infty$, one finds that at $s \to 0$ the correlator $\tilde{Z}(s)$ must behave as $\tilde{Z}(s) \approx As^{-\mu}$, $\mu < 1$, i.e., $Z(t) \approx At^{\mu-1}/\Gamma(\mu)$, where $0 < \mu < 1$ and $\Gamma$ is the Gamma function. The Einstein diffusion corresponds to $\mu = 1/2$. It is important to note that the correct solution (18) (and consequently the VAF (11) and MSD (13)) can be obtained without an explicit use of the correlation properties of the thermal noise [24]. Then, the knowledge of the exact solution allows one to determine which should be the properties of the driving noise that lead to the required result.

Consider for a moment the generalized Langevin equation in its most common form [18] with the velocity in the memory integral instead of acceleration,

$$M\dot{v}(t) + \int_0^t \gamma(t-t')v(t')dt' = \xi(t). \quad (19)$$

Here $M$ is the particle mass, $\xi(t)$ is the random force, and $\gamma(t)$ a new memory kernel. Multiplying this equation by $\xi(0) = M\dot{v}(0)$ and averaging, with the use of the relations for the stationary process, $\langle \dot{v}(0)v(t) \rangle = -\langle \dot{v}(t)v(0) \rangle$, $\langle \dot{v}(0)v(0) \rangle = 0$, Eq. (19) can be rewritten as

$$H(t) = \langle \xi(t)\xi(0) \rangle$$
$$= -M^2 \frac{d^2}{dt^2}\langle v(t)v(0) \rangle - M\int_0^t \gamma(t-t') \frac{d}{dt'}\langle v(t')v(0) \rangle dt'. \quad (20)$$

Using $\ddot{F}(t) \div s^2\tilde{F}(s) - sF(+0) - \dot{F}(+0)$ and $\dot{F}(t) \div s\tilde{F}(s) - F(+0)$, we obtain

$$\tilde{H}(s) = -M\left[s\tilde{\phi}(s) - \phi(0)\right]\left[Ms + \tilde{\gamma}(s)\right] = M\phi(0)\tilde{\gamma}(s). \quad (21)$$

This is the well-known relation between the memory kernel $\gamma$ and the correlator of the random force, $H(t) = k_BT\gamma(t)$ (the so-called second FDT [18]).

In the same way we obtain for the correlator $N(t) = \langle \zeta(t)\zeta(0) \rangle$ in our problem of the hydrodynamic Brownian motion,

$$\tilde{N}(s) = \gamma^2\tilde{\phi}(s) - \left[M^2s + Ms\tilde{\Gamma}(s) - \gamma\tilde{\Gamma}(s)\right]\left[s\tilde{\phi}(s) - \phi(0)\right]. \quad (22)$$

The Einstein diffusion at long times is reached if Eq. (18) holds, when

$$\tilde{N}(s)/\phi(0) = \gamma M + \tilde{\Gamma}(s)\left[Ms - \gamma\right] \quad (23)$$

and

$$N(t) = k_BT\gamma\left[\delta(t) - \frac{1}{\tau}\sqrt{\frac{\tau_R}{\pi t}}\theta(t) - \frac{1}{2}\sqrt{\frac{\tau_R}{\pi t^3}}\theta(t)\right], \quad (24)$$

where $\theta(t)$ is the Heaviside function. This expression is similar to that found in Ref. [17], except for the second term in square brackets that is missing there.

Now, let us consider the studied problem from another view. If we average Eq. (19) with a regular external force $F(t)$,

$$M\dot{\upsilon}(t) + \int_0^t \gamma(t-t')\upsilon(t')\mathrm{d}t' = \xi(t) + F(t), \quad (25)$$

using $\langle \upsilon(0) \rangle = \langle \xi(t) \rangle = 0$ and the Laplace transformation (denoted by the index $s$, e.g., the transform of $\langle \upsilon(t) \rangle$ will be $\langle \upsilon(t) \rangle_s$), we find

$$\langle \upsilon(t) \rangle_s = \frac{\langle F(s) \rangle_s}{Ms + \tilde{\gamma}(s)}. \quad (26)$$

From here the admittance (or mobility of the particle) [18] is

$$\tilde{\mu}(s) = \frac{1}{Ms + \tilde{\gamma}(s)}. \quad (27)$$

Since the solution of (25) for a free particle is

$$\tilde{\phi}(s) = \frac{k_B T}{Ms + \tilde{\gamma}(s)}, \quad (28)$$

we find

$$\tilde{\phi}(s) = k_B T \tilde{\mu}(s). \quad (29)$$

This relation is called the basic theorem or the generalization of the first FDT [18]. It also holds

$$\langle \upsilon(t) \rangle_s = \tilde{\mu}(s)\langle F(s) \rangle_s = \frac{1}{k_B T}\tilde{\phi}(s)\langle F(s) \rangle_s. \quad (30)$$

This equation looks like the one used in Ref. [12] (Eq. 3b),

$$\upsilon(t) = \frac{1}{k_B T}\int_0^t \phi(t-t')F(t')\mathrm{d}t', \quad (31)$$

but in [12] $F$ is the random force assumed to be zero at $t < 0$, and $\upsilon$ is the instantaneous velocity ($\upsilon(0) = 0$). However, Eq. (30) is valid for the mean velocity and the external force $F$. When the regular external force is absent, the mean velocity is zero and Eq. (30) cannot be used. The derivation of the VAF in [12] is thus flawed. Nevertheless, the basic theorem (29) allows us to find the correlators of the random force in the Langevin equation and the velocity. If Eq. (19) is multiplied by $\upsilon(0)$ and averaged, in the Laplace transform we obtain for the VAF:

$$\tilde{\phi}(s) = \frac{M\phi(0) + \tilde{Z}(s)}{Ms + \tilde{\gamma}(s)}, \quad (32)$$

where the transformed correlator $Z(t) = \langle \xi(t)\upsilon(0) \rangle$ reads

$$\tilde{Z}(s) = \tilde{\phi}(s)\left[Ms + \tilde{\gamma}(s)\right] - M\phi(0). \quad (33)$$

If the relation $\tilde{\phi}(s) = k_B T \tilde{\mu}(s)$ holds, $Z(t) = 0$. If we require that normal diffusion takes place at long times and there is equipartition at $t = 0$, again $Z(t) = 0$ at $t > 0$. However, in the hydrodynamic Brownian motion the condition $Z(t) = 0$ leads to super-diffusion. We have shown above for Eq. (2) that $Z(t) = \langle \zeta(t) \upsilon(0) \rangle$ must differ from zero. This is consistent with the general relation (29): for $\tilde{\phi}(s)$ we have the equation

$$\tilde{\phi}(s) = \tilde{\mu}(s)\left\{\left[M + \tilde{\Gamma}(s)\right]\phi(0) + \tilde{Z}(s)\right\}, \quad (34)$$

which together with (29) gives

$$\tilde{Z}(s) = -\phi(0)\tilde{\Gamma}(s). \quad (35)$$

This is the same relation that we have obtained from the requirement that Einstein's diffusion takes place at long times.

Note that the "fundamental hypothesis" $Z(t) = 0$ is used in the cited works [16, 17] also for the hydrodynamic Brownian motion. Then the derivation of the VAF in [17] is based on the solution of Kubo's equation (19) by multiplying it with $\upsilon(0)$ and averaging. However, evaluating the integral in Eq. (2) directly *per partes* to obtain Eq. (19), the infinity appears at $t' = t$ since $\Gamma(t - t') \sim (t - t')^{-1/2}$. Nevertheless, Eq. (19) can really be used to solve the problem of the hydrodynamic Brownian motion. However, then the noise force $\xi(t)$ will not correspond to the force $\zeta(t)$ from Eq. (2) (or the true thermal force $\eta(t)$ from (15)). It is seen from the correlators of the random forces with the velocity $\upsilon(0)$. While Eq. (19) possesses the correct VAF (for $\gamma(t) = -\gamma(\tau_R/4\pi)^{1/2}t^{-3/2}$, $t > 0$ [17]), it is possible only when the relation $\langle \xi(t)\upsilon(0)\rangle = 0$ is used. Using the correlator $\langle \xi(t)\upsilon(0)\rangle = (-k_B T\gamma/M)(\tau_R/\pi t)^{1/2}$ and the correct solution for the VAF, the correspondence between the hydrodynamic Langevin equation (2) and Kubo's equation (19) can be established for $\gamma(t) = -(\gamma/\tau)(\tau_R/\pi t)^{1/2}$, $t > 0$, that significantly differs from $\gamma(t)$ found in [17]. Thus, to obtain the correct solution for the hydrodynamic Brownian motion from Kubo's equation with the same properties of the thermal force as in the original hydrodynamic Langevin equation, one must accept that the correlator $\langle \xi(t)\upsilon(0)\rangle$ is nonzero.

Finally, let us return to the question of the "color" of thermal noise. In Ref. [10] it was experimentally probed for particles trapped in a harmonic potential. Ignoring the particle inertia at low frequencies (long times), the trapping force dominates over friction. Then the Langevin equation (2) with the force $F = -Kx(t)$, where $K$ is the stiffness of the harmonic potential, is reduced to $Kx(t) \approx \zeta(t)$. From here Franosch et al. [10] used $\langle \zeta(t)\zeta(0)\rangle \approx K^2\langle x(t)x(0)\rangle$ and determined

$$\langle \zeta(t)\zeta(0)\rangle = -\frac{1}{2}k_B T\gamma\left(\frac{\tau_R}{\pi t^3}\right)^{1/2}. \quad (36)$$

It is evident that this is not correct since such a derivation requires that also $Kx(0) \approx \zeta(0)$ holds, which is not true: as opposite, at $t \to 0$ the force term $Kx(t)$ in Eq. (2) is less important than the inertia term, the memory integral, and the frictional force $\gamma\upsilon$. One more shortcoming in Ref. [10] is of the same kind as in [17] and consists in the above discussed use of Kubo's equation (19). Here, from Eq. (24), a different correlator $\langle \zeta(t)\zeta(0)\rangle$ has been obtained. For $t > 0$

$$\langle \zeta(t)\zeta(0)\rangle = -k_B T\gamma \left(\frac{\tau_R}{\pi t}\right)^{1/2} \left(\frac{1}{\tau}+\frac{1}{2t}\right). \qquad (37)$$

This correlation function is the same as that for the true thermal noise $\eta(t)$ since $\langle \zeta(t)\zeta(0)\rangle = \langle \eta(t)\eta(0)\rangle + C$ (see the relation between $\zeta$ and $\eta$ below Eq. (15)). The time-independent term $C$ is zero since at $t \to \infty$ both the correlators converge to zero. The time correlation function of the thermal noise in incompressible fluids thus at long times approaches zero as $t^{-1/2}$ instead of $t^{-3/2}$ found in [10]. Since within the linear theory the properties of the thermal force do not depend on the external force [18], this result, obtained from the Langevin equation for a free particle, will be the same as for trapped particles studied in Ref. [10].

## 4 Conclusion

This work deals with statistical properties of the thermal noise driving the Brownian particles in incompressible fluids. Their motion is described by the Langevin equation with the random force $\zeta(t)$ and the Boussinesq force that takes into account the added mass effect and is related to the history of the particle motion. If the initial time is $t = 0$, the latter force contains an additional term to the Stokes drag, which is expressed as a convolution of the particle acceleration and the hydrodynamic memory kernel $\sim t^{-1/2}$. Using an effective method of solving linear generalized Langevin equations [21, 24], we have shown that the expected Einstein diffusion follows from such a hydrodynamic Langevin equation only in the case when the random force $\zeta(t)$ at $t > 0$ correlates with the velocity of the particle at time $t = 0$, whereas for the true thermal force $\eta(t)$ we have $\langle \eta(t)\upsilon(0)\rangle = 0$ due to causality. This is in contradiction with the "fundamental hypothesis" (according to which $\zeta(t)$ and $\upsilon(0)$ are uncorrelated) applied in a number of papers dealing with the normal and generalized Langevin equation.

An analytical expression for the time correlation function $\langle \zeta(t)\upsilon(0)\rangle$ has been found. It is in accordance with the exact solutions obtained without the explicit use of the correlation properties of the random force [24, 25]. We have also shown that the correct mean square displacement of the particles exhibiting normal diffusion at long times can be obtained coming from another basic theorem from the linear response theory; this theorem joins the mobility of the particle and its velocity autocorrelation function.

Finally, we addressed the question of the correlation properties of the thermal noise itself, recently probed in the experiments [10]. We have shown that the interpretation of these experiments should be corrected and calculated the time autocorrelation function of the thermal noise in incompressible fluids, $\langle \eta(t)\eta(0)\rangle = \langle \zeta(t)\zeta(0)\rangle$. The difference between our results and those found in previous works [10, 17] is significant. The thermal force can be observed in experiments, for instance, in electric circuits as the thermal voltage across a resistor [18, 26]. A detailed discussion how the thermal noise (its color) could be directly measured in fluids by the separation of the noise from the Brownian particle´s response to it is given in Ref. [27]. The proposed method assumes the use of optical tweezers that was realized in the discussed paper [10] (see also [28]). The main result of our paper, Eq. (37), can be used for the interpretation of such experiments as follows: at long times ($t \gg \tau_R$) the correlator of the thermal force could be determined from measurements of the positions of Brownian particles in a harmonic trap according to the formula

$$\langle \eta(t)\eta(0) \rangle \approx K^2(1 + 2t/\tau)\langle x(t)x(0) \rangle, \quad (38)$$

where $K$ is the force constant of the external potential. This formula follows from Eq. (37) and the long-time result for the positional autocorrelation function of a Brownian particle in the trap [10, 14, 15], $\langle x(t)x(0) \rangle = k_B T/K - X(t)/2 \approx -k_B T\gamma(\tau_R/4\pi t^3)^{1/2}/K^2$.

The time correlation function for the thermal noise obtained in [10] (Eq. (36)) and that in the present work differ, although they come from the same particle positional function measured in experiments. The difference is due to a different interpretation of the data. In our approach the "collor" of the noise follows from the function $\langle x(t)x(0) \rangle$ calculated from the hydrodynamic Langevin equation (15). In the papers [10, 16, 17] the generalized Langevin equation (19) is used. While Eq. (19) gives the correct solution for the mean square displacement of the particle, this solution is reached for the random force which has properties different from those of the true thermal noise present in Eq. (15). One can see it from the correlation function $\langle \zeta(t)v(0) \rangle$ that is zero for Kubo´s equation. For the hydrodynamic Langevin equation it is nonzero, as shown in Section 3. Thus, the different interpretations lead to different colors of the thermal noise. Since in the linear approximation the thermal force is not affected by the external forces, our results can be used in consistent solution of the problem of the Brownian motion in harmonic traps, which is related to numerous experiments on colloidal systems.

**Acknowledgement**  This work was financially supported by the Agency for the Structural Funds of the EU within the projects NFP 26220120021, 26220120033, and by the grant VEGA 1/0370/12.